\documentclass[intlimits,twoside,a4paper]{article}
\usepackage{amsmath,amssymb}
\usepackage{graphicx}
\usepackage[T2A]{fontenc}
\usepackage[cp1251]{inputenc}

\usepackage{color}

\usepackage[eqsecnum]{cmpj2}

\issue{2016}{19}{1}{13609}
\doinumber{10.5488/CMP.19.13609}

\title[Flow of oligomeric fluid]
{Pressure-driven flow of oligomeric fluid in nano-channel with complex structure.
A dissipative particle dynamics study\thanks{It is our pleasure to dedicate this paper to Professor Stefan Soko{\l}owski,
our Colleague and Mentor for many years.\protect}}
\author[J.M. Ilnytskyi, P. Bryk, A. Patrykiejew]{J.M. Ilnytskyi\refaddr{label1}, P. Bryk\refaddr{label2}, A. Patrykiejew\refaddr{label2}}
\addresses{
\addr{label1} Institute for Condensed Matter Physics of the National Academy of Sciences of Ukraine, \\ 1~Svientsitskii St., 79011 Lviv, Ukraine
\addr{label2} Department for the Modeling of Physico-Chemical Processes, Maria Curie-Sk{\l}odowska University, \\ 20--031 Lublin, Poland
}

\date{Received November 2, 2015, in final form January 26, 2016}
\authorcopyright{J.M. Ilnytskyi, P. Bryk, A. Patrykiejew, 2016}

\newcommand{\vvec}[1]{\mathrm{\bf #1}}
\newcommand{\vhat}[1]{\hat{\mathrm{\bf #1}}}
\newcommand{\idx}[1]{_{\mathrm{#1}}}
\newcommand{\upx}[1]{^{\mathrm{#1}}}

\begin{document}
\maketitle
\begin{abstract}
We develop a simulational methodology allowing for simulation of the pressure-driven flow
in the pore with flat and polymer-modified walls. Our approach is based on dissipative
particle dynamics and we combine earlier ideas of fluid-like walls and reverse flow.
As a test case we consider the oligomer flow through the pore with flat walls
and demonstrate good thermostatting qualities of the proposed method.
We found the inhomogeneities in both oligomer shape and alignment across the pore
leading to a non-parabolic velocity profiles.
The method is subsequently applied to a nano-channel decorated with a polymer brush
stripes arranged perpendicularly to the flow direction. At certain threshold value of
a flow force we find a pillar-to-lamellar morphological transition, which leads to
the brush enveloping the pore wall by a relatively smooth layer. At higher flow rates,
the flow of oligomer has similar properties as in the case of flat walls, but for the
narrower effective pore size. We observe stretching and aligning of the polymer molecules
along the flow near the pore walls.
\keywords Pouiseuille flow, polymer brush, oligomers, dissipative particle dynamics
\pacs 68.03.Cd, 68.08.-p, 68.03.-g, 68.35.Md, 68.47.Mn
\end{abstract}

\section{\label{I}Introduction}

Understanding the behavior of polymers attached to surfaces is of importance in many research areas including
biophysics, polymer-induced effective interactions in colloidal suspensions, chromatographic separation,
catalysis, and drug delivery \cite{Zhao2000}. Grafting polymer chains can significantly
alter the properties of the surface and make it, for example, biocompatibile or
responsive to external stimuli \cite{Minko2006}. Due to the large field of
potential applications, polymer brushes have been the subject of many theoretical studies.
In the seminal works Alexander \cite{Alexander1977}, de Gennes \cite{deGennes1976,deGennes1980}
have calculated the brush profile and explored analytically the impact of grafting density and molecular weight.
Since then, the properties of tethered chains have been investigated by means of
self-consistent field theory \cite{Milner1988,Scheutjens1986,Wijmans1992,Netz1998,Carigano1995},
polymer density functional theory \cite{McCoy2002,Cao2006,Xu2009,Xu2010,Milchev2010,Borowko2013},
and computer simulation \cite{Grest1996,Weinhold1994,Pastorino2006,MacDowell2006,Fouqueau2007}.
Many theoretical predictions have been confirmed by experiment
\cite{Hadziioannou1986,Taunton1988,Taunton1990,Auroy1991}.

Polymer brushes can be used to tailor static properties of surfaces,
such as wettability, as well as dynamical, such as hydrodynamic boundary conditions and friction.
Fluid flow in polymer grafted nano\-pores can be described via continuum
hydrodynamic equations (e.g., the Brinkman equation \cite{Brinkman1947}) with a priori assumed permeability related
to the monomer density profile. The resulting velocity profile is sensitive to the assumed form
of the monomer profile \cite{Milner1991}. However, the continuum hydrodynamic description of a flow has not
been firmly established on the nanoscale \cite{Mueller2008}.
This is important in the context of micro- and nanofluidic devices \cite{Squires2005}.
Downsizing a channel to the nanoscale, increases the surface-to-volume ratio and
introduces new physical phenomena not observed in the macroscale \cite{Schoch2008}.
Covering the surface by a polymer brush may introduce a pronounced reduction of friction, which
lowers the pressure difference required to maintain the flow through a nanochannel \cite{Klein1991}.
Flow in polymer brushes has been the subject of numerous simulational studies
in recent years \cite{Lai1993,Doyle78,Grest1996b,Adiga2005,Huang2006,Masoud2011,Dimitrov2008}.

Recently, the equilibrium properties of binary mixture confined in a slit-like pore decorated
with polymer brush stripes were studied by means of
dissipative particle dynamics (DPD) \cite{ISP1,ISP2}.
It was found that, depending on the geometrical parameters characterizing the system
(the size of the pore and the width of the stripes), several different structures (or morphologies)
inside the pore can be formed. Such patterned brushes can be fabricated experimentally by means of electron
beam litography \cite{Paik2010}. In the present paper, we wish to study nonequilibrium properties
of such system by considering the pressure driven oligomer flow inside a channel with either flat or brush-modified walls.
In particular, we focus on three features such as: (i) the microstructure of a flow depending on
the molecular mass of an oligomer and the magnitude of a bulk flow force; (ii) flow-induced morphology changes;
and (iii) the effect of the patterned brush decoration of the walls on the properties of the flow.
Our paper is arranged as follows:
In section~\ref{II} we introduce a new simulation method which combines the ideas of fluid-like
walls and reverse flow to minimise the near-wall artefacts and maintain constant
temperature under flow condition. As a simple test, we apply the method to the case
of oligomer flow through the pore with flat walls.
In section~\ref{III}, the analysis is extended to the case when the walls are
modified by a polymer brush arranged in a form of stripes. Conclusions are provided in section~\ref{IV}.

\section{\label{II}Flow of oligomeric fluid through a channel with flat walls}

Let us first consider the simulational approach employed in this study.
We use the non-equilibrium extension of the DPD technique in a form discussed
by Groot and Warren \cite{GrWarr}. This is a mesoscopic method that operates
at a level of coarse-grained beads, each representing either a fragment
of a polymer chain or a collection of solvent particles. The force acting
on $i$th bead due to its pairwise interaction with $j$th bead can be written
as
\begin{equation}
  \vvec{F}_{ij} = \vvec{F}\upx{C}_{ij} + \vvec{F}\upx{D}_{ij}
  + \vvec{F}\upx{R}_{ij}\,,
\end{equation}
where $\vvec{F}\upx{C}_{ij}$, $\vvec{F}\upx{D}_{ij}$ and
$\vvec{F}\upx{R}_{ij}$ denote the conservative, dissipative and
random contribution, respectively. These have the following form \cite{GrWarr}
\begin{equation}\label{FC}
  \vvec{F}\upx{C}_{ij} =
     \left\{
     \begin{array}{ll}
        a(1-r_{ij})\vhat{r}_{ij}, & \qquad r_{ij}<1,\\
        0,                       & \qquad r_{ij}\geqslant 1,
     \end{array}
     \right.
\end{equation}
\begin{equation}\label{FD}
  \vvec{F}\upx{D}_{ij} = -\gamma
  w\upx{D}(r_{ij})(\vhat{r}_{ij}\cdot\vvec{v}_{ij})\vhat{r}_{ij},
\end{equation}
\begin{equation}\label{FR}
  \vvec{F}\upx{R}_{ij} = \sigma
  w\upx{R}(r_{ij})\theta_{ij}\Delta t^{-1/2}\vhat{r}_{ij}.
\end{equation}
Here, $\vvec{v}_{ij}=\vvec{v}_i-\vvec{v}_j$, $\vvec{v}_i$ and
$\vvec{v}_j$ are the velocities of the beads, $\theta_{ij}$ is
Gaussian random variable, $\langle \theta_{ij}(t) \rangle=0$, $\langle
\theta_{ij}(t) \theta_{kl}(t')\rangle=(\delta_{ik}\delta_{il} +
\delta_{il}\delta_{jk})\delta(t-t')$ and $\Delta t$ is the
time-step of the integrator. As already discussed in references~\cite{Mueller2008,Pastorino2009},
the effective range of friction between beads can be modified by adjusting the shape of
the weight functions $w\upx{D}(r_{ij})$ and $w\upx{R}(r_{ij})$. We use
the following general form for $w\upx{R}(r_{ij})$:
\begin{equation}
w\upx{R}(r_{ij})=
  \left\{
  \begin{array}{ll}
     (1-r_{ij})^\beta, & \qquad r_{ij}<1,\\
     0,            & \qquad r_{ij}\geqslant 1,
  \end{array}
  \right.
\end{equation}
where the exponent $\beta$ is adjusted, and the weight function $w\upx{D}(r_{ij})$
is set equal to $[w\upx{R}(r_{ij})]^2$ according to Espa\~{n}ol and Warren \cite{EspWarren_1995}
arguments. Likewise, it is required that $\sigma^2=2\gamma$.

The oligomers and tethered polymer chains (if any) are represented as necklaces of
beads bonded together via harmonic springs, the force acting on $i$th bead
from the interaction with its bonded neighbour, $j$th bead, is
\begin{equation}\label{FB}
  \vvec{F}\upx{B}_{ij} = -kr_{ij}\vhat{r}_{ij}\,,
\end{equation}
where $r_{ij}=|\vvec{r}_{ij}|$, $\vvec{r}_{ij}=\vvec{r}_i-\vvec{r}_j$
is the vector connecting the centers of $i$-th and $j$-th beads,
$\vhat{r}_{ij}=\vvec{r}_{ij}/r_{ij}$ and $k$ is the spring constant.
The same bonding force is used to tether the end polymer bead to the
surface. The length, mass, time and energy (expressed via
$T^*\equiv k\idx{B}T$) units are all normally set equal to unity.

Let us now turn to the case where the fluid (or a mixture of fluids) is confined
within a slit-like pore. In order to commence a simulation of the pressure-driven flow,
it is required to provide a set of rules defining the behaviour of the fluid particles at walls,
and a prescription for the construction of the walls.
These rules should recover the well known cases of hydrodynamic flow such as
the Poiseuille flow (i.e., a flow of a Newtonian fluid with no-slip boundary conditions and
a parabolic velocity profile). On the other hand, for the flow of a polymeric fluid
(i.e., a non-Newtonian fluid) the set of rules should lead to the slip boundary conditions.
The simplest set of rules comprise elastic reflections off the
wall \cite{Ashurst_refl-a,Ashurst_refl-b}. Unfortunately, they give rise to a hydrodynamic slip for Newtonian fluids,
as well as  suffer from near-wall density artifacts at higher density. This can be traced back to the fact
that the atoms repelled each other strongly but did not interact with the wall
until they attempted to cross \cite{Ashurst_75}.

A number of more sophisticated set of rules have been suggested. One option
is to form the crystalline walls of a few layers of frozen (or having large mass)
particles \cite{Koplik_88,dpd8-a,dpd8-b,Visser-2005,frozen-walls-a,frozen-walls-b,frozen-walls-c}.
The interaction between the bulk fluid particles and those of the wall
creates the near-wall drag which leads to the formation of the Poiseuille flow.
A drawback of this approach is the propagation of the crystalline order into the
near-wall regions of bulk fluid. This effect is perfectly physical for the atomic
molecular dynamics simulation, where the solid wall mimics a real crystalline structure.
However, for the mesoscopic DPD simulations, each soft bead is assumed to
represent a meso-scale portion of the material, on which scale the atomic
crystalline structure is smeared-out.

This drawback can be avoided by using the structureless fluid-like walls \cite{Ashurst_75}.
The walls in this case are made of
the fluid confined in the slabs adjacent to the pore boundary, and the elastic reflections
are applied on both sides of the boundary. Therefore, bulk and wall fluid
particles are immiscible. Still, the interaction between near-wall
beads on the opposite sides of the boundary creates a near-wall drag ensuring
no-slip boundary condition for Newtonian fluids.

Another important issue in flow simulation is to avoid the system
overheating due to the presence of the body force. This problem was addressed
in several studies, cf. for example references~\cite{Ghosh,Past_07}.
In the molecular dynamics simulation, the excessive energy is absorbed by an external thermostat, in either
bulk or near-wall form \cite{Ghosh}. In DPD simulations, the thermostat is ``internal'',
provided by the balance between interparticle friction and random forces.
For the case of a flow, some means for dissipation of additional energy related
to the body force should be provided.
One of the elegant ways to do this is the concept of a reverse
flow \cite{Muller-Plathe_99,Backer_05,Fedosov2010}. In this approach,
the simulation box contains two sub-flows driven oppositely. The total
force applied to the system is equal to zero and a no-slip boundary
is formed at the interface between two opposite flows of Newtonian fluids.

In our study, we combine both concepts by employing the fluid-like walls on both
boundaries of a pore and initiating contraflows (reverse flows) within them. Separation between
the main pore and contraflow-containing walls prevents intermixing between the beads
from both regions. This is important both in the case when the flow of a mixture
is considered, or in the case of polymer modified walls, where polymer chains are
tethered to the boundary between the main pore and fluid-like wall.
However, the existence of the reflective boundaries does not prevent a friction
between the beads located on the opposite sides of the boundary, enabling the formation of the no-slip boundary condition for Newtonian fluids.

\begin{figure}[!t]
\centering
\includegraphics[width=8cm]{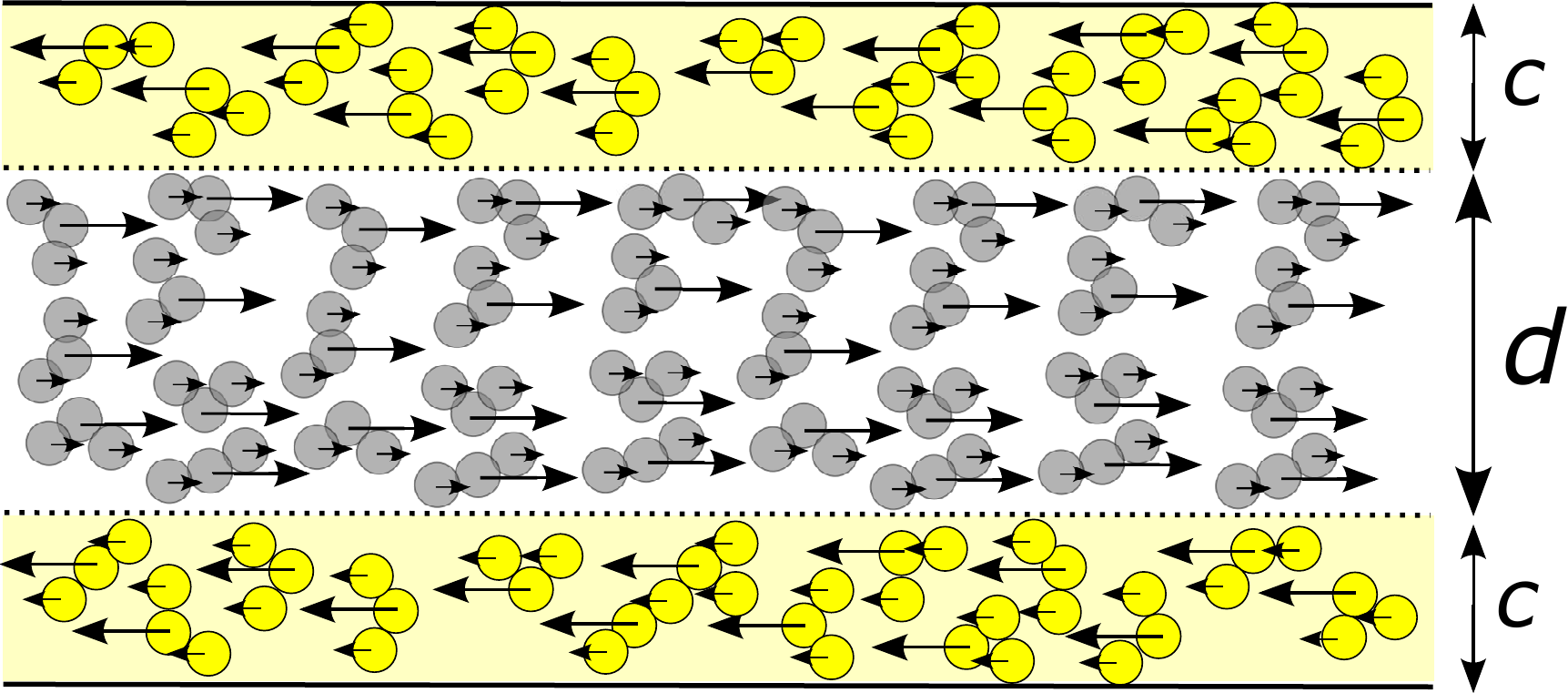}
\caption{\label{geom_nobr_olig} (Color online) Simulation box geometry showing
the pressure-driven flow of the oligomer of $L\idx{o}=3$ beads in a
slit-like pore of size $d$ (gray beads). Contraflow containing
an oligomer of the same length is contained in two fluid-like
walls of size $c$ (yellow beads). The direction of the applied flow
force is shown via arrows of different length reflecting the  force
distribution biased towards the middle bead, see text for details.}
\end{figure}

In this section, we consider the pressure-driven flow of oligomeric one-component
fluid through the pore with flat walls. The oligomers of length $L\idx{o}=1,4,10$ and
$20$ beads are considered. The aim is twofold. Firstly, we would like to test to what
typical values of bulk force the approach outlined above can be stretched without
violation of temperature conservation. Secondly, we aim to study the flow microstructure
depending on molecular length of the flowing oligomer and the magnitude of a flow force.
The geometry of the system
is illustrated in figure~\ref{geom_nobr_olig}. Here, $X$-axis runs from left to right,
$Z$-axis~--- from bottom to top, $Y$-axis coincides with the viewing direction. The
simulation box is of dimensions $L_x=80$, $L_y=50$ and $L_z=26.667$ with the periodic
boundary conditions applied along $X$ and $Y$ axes, the pore size is $d=13.333$, the
size of the contraflow regions is $c=d/2=6.667$. The chains in contraflow regions are of
the same length $L\idx{o}$ as in the main pore. Therefore, the total number of main
and contraflow chains is the same. All beads are assumed to be of the same type,
which is reflected in the fact that the parameter $a$ in equation (\ref{FC}) that controls
the bead repulsion is the same for all pairwise interactions, $a=25$.
The pore and the fluid-like walls are separated via
the internal boundaries (shown as dashed lines in figure~\ref{geom_nobr_olig}),
impenetrable for the beads on both sides by applying the reflection algorithm
described in detail in reference~\cite{ISP1}. The same reflection algorithm is used at the external walls (solid lines in the same figure) but, alternatively, the periodic boundary conditions can be used in $Z$ direction, similarly to the original reverse
flow setup \cite{Muller-Plathe_99,Backer_05,Fedosov2010}.

\begin{figure}[!t]
\centering
\includegraphics[width=8cm]{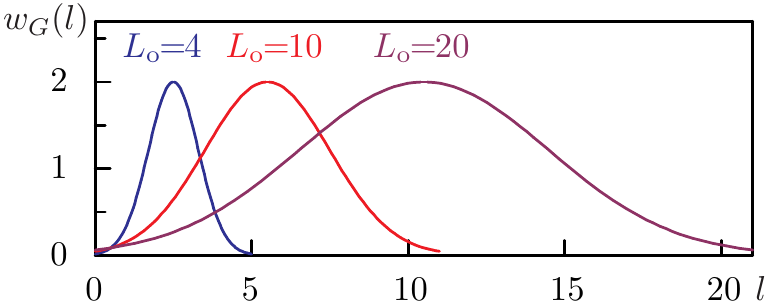}
\caption{\label{gauss}  (Color online) Weight function $w_\text{G}(l)$ for the amount of bulk force
applied to $l$th bead of the oligomer of length $L\idx{o}$, equation~(\ref{Gauss_distr}).}
\end{figure}

Each $i$th bead within a pore is subjected to the flow force of certain amount $f_i$
\begin{equation}\label{FF}
  \vvec{F}\upx{FL}_i = f_i\vhat{x},\hspace{3em}\vhat{x}=\{1,0,0\},
\end{equation}
applied along $X$-axis, where $f_i>0$, this is indicated by the right-hand side directed
arrows in figure~\ref{geom_nobr_olig}. The beads in the contraflow regions are
subjected to the force $\vvec{F}\upx{FL}_i=-f_i\vhat{x}$, indicated as
reversely directed arrows in the same figure. Several options are available for choosing
the amount of $f_i$. The simplest one would be to choose $f_i\equiv f$, the
same amount for each bead.
However, such an algorithm could lead to less than optimal match of the
micro-fluctuations of the applied pressure in real systems, since the polymer molecules
tend to form coils with varying distribution of the density.
Another, rather extreme option would
be to apply the amount $fL\idx{o}$ to the middle bead only. The other beads  feel
this force indirectly and are delayed via the elastic spring forces. The latter approach might suffer
from large fluctuations of bond lengths and slower relaxation of the intra-chain vibrations,
due to the soft nature of the model. In our view, a reasonable compromise can be achieved
by applying a fixed amount of the force $fL\idx{o}$ to each oligomer, but biasing
it towards the middle bead of the chain. Namely, assuming that the beads of an oligomer
are numbered sequentially as $l=1,\ldots, L\idx{o}$, then the amount of the force applied
to the bead number $l$ is found according to the Gaussian distribution:
\begin{equation}\label{Gauss_distr}
f(l) = f w_\text{G}(l),\hspace{3em} w_\text{G}(l)=2\exp\left[-\frac{(l-\bar{l})^2}{\sigma^2}\right].
\end{equation}
Here, $\bar{l}=(L\idx{o}+1)/2$ is the mid-index of the chain, and the
breadth of the distribution is given by $\sigma=L\idx{o}/(2\sqrt{\pi})$.
The distribution is normalized to $L\idx{o}$:
\begin{equation}\label{gauss_norm}
\sum_{l=1}^{N\idx{o}}w_\text{G}(l)=\int_{-\infty}^{+\infty}w_\text{G}(l)\rd l=L\idx{o}.
\end{equation}
The shape of the weight function $w_\text{G}(l)$ is shown in figure~\ref{gauss} for
the cases of $L\idx{o}=4$, $10$ and $20$. As a result, the total force applied
to the oligomer of $L\idx{o}$ beads is equal to $fL\idx{o}$, but it is biased
towards the middle beads (illustrated by arrows of different length in
figure~\ref{geom_nobr_olig}).

\begin{figure}[!b]
\centering
\includegraphics[width=0.95\textwidth]{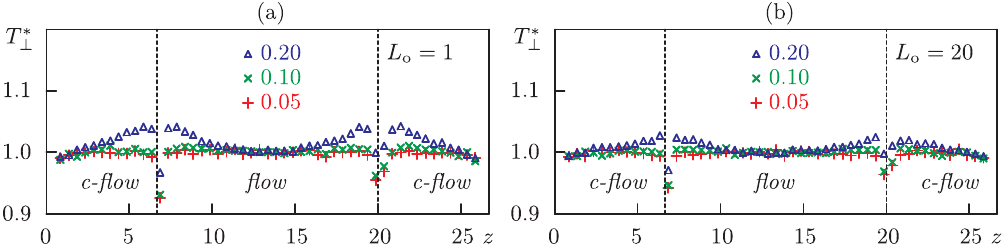}
\caption{\label{t_conserv}  (Color online) $T_{\perp}^*$ profile at various amplitudes of bulk
force $f$ indicated in the figure. (a) simple liquid, $L\idx{o}=1$;
(b) longest oligomer, $L\idx{o}=20$.}
\end{figure}

The acceleration of the fluid beads due to applying the bulk force
affects the accuracy of the integrator, as far as the expression for the
coordinates at the time instance $t+\Delta t$ contains the term proportional
to $v(t)\Delta t$, where $v(t)$ is the velocity of the particle at the time
instance $t$. The only way to keep the same numerical accuracy of the
integrator is to reduce the time-step $\Delta t$ of the integrator. In our
simulations, we use the value $\Delta t=0.001$, about $40$ times smaller
compared to the values typically used in the case of equilibrium simulation.
Temperature conservation is one of the most important indicators of the
accuracy of the integrator. Following reference~\cite{Past_07}, we consider
the transverse temperature, which is evaluated from the two components of
the velocity perpendicular to the flow direction:
$T_{\perp}^*=m\langle v_y^2\rangle/2+m\langle v_z^2\rangle /2$. The profiles of
$T_{\perp}^*$ for two extreme cases of oligomer length $L\idx{o}=1$ (simple fluid)
and $20$ with respect to $z$ coordinate are shown in figure~\ref{t_conserv}
(a) and (b), respectively. We allow the maximum deviation of these profiles from
the required value $1$ not to exceed $3-4\%$. As follows from figure~\ref{t_conserv},
this is achieved for all $L\idx{o}=1-20$ if the flow force magnitude is restricted to $f\leqslant 0.2$.
At larger values, $f > 0.2$, the system is prone to local heating near the internal walls,
which signals a  breakdown of this thermostatting method. We should also remark that
for the setup with no contraflow regions, no thermostatting can be achieved at all:
the temperature was found to rise monotonously even for the smallest considered
values of~$f$.

\begin{figure}[!t]
\centering
\includegraphics[width=0.95\textwidth]{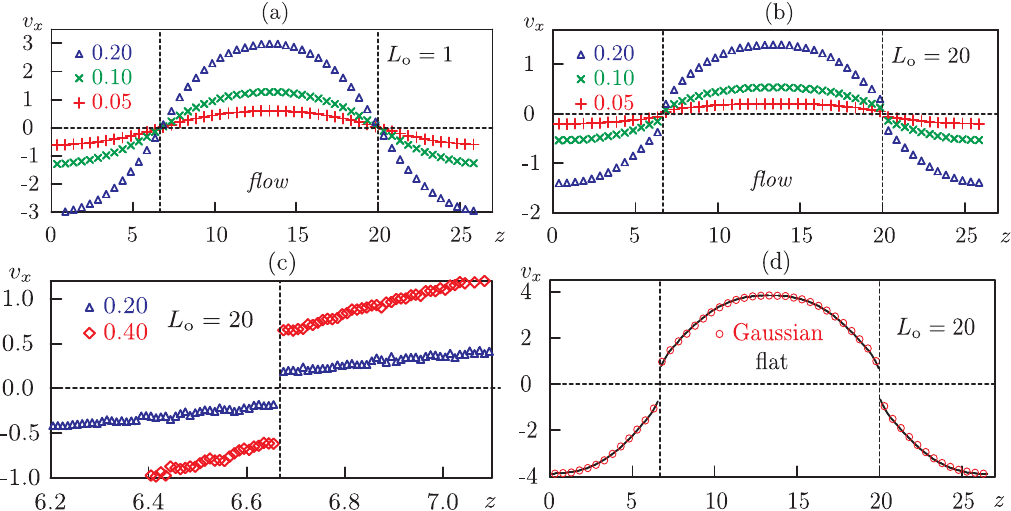}
\caption{\label{vx_fit}  (Color online) Profiles for the velocity component along the flow direction, $v_x$,
evaluated bead-wise at several values of $f$ (indicated in the figure).
(a) the case of simple fluid, $L\idx{o}=1$; (b) the case of the longest oligomer considered, $L\idx{o}=20$;
(c) near-wall behaviour of the velocity profiles for $L\idx{o}=20$; (d) the velocity profiles for $f=0.4$
and for $L\idx{o}=20$ evaluated with the Gaussian distribution of the bulk force, [cf. equation~(\ref{Gauss_distr})--(\ref{gauss_norm})],
(symbols), and a uniform distribution of the bulk force (solid line).}
\end{figure}

The profiles for the velocity components $v_x$ of individual beads along the flow direction
are built by binning the pore along the $Z$-axis. These are shown in figure~\ref{vx_fit} for the cases of $L\idx{o}=1$ and $L\idx{o}=20$ oligomer length obtained at various flow force
amplitudes $f=0.05, 0.1$ and $0.2$. For the case of simple fluid (a), almost
perfect parabolic shape is achieved inside the flow region indicating the properties
of a Newtonian fluid. The velocity drops to zero exactly at the pore walls giving rise to the no-slip
boundary condition. In this case, the Stokes formula can be used to estimate the
viscosity of the fluid. With an increase of the oligomer length, $L\idx{o}$, the shape of
the velocity profile gradually diverges from a parabolic one and turns into a bell-like shape
at $L\idx{o}=20$, as seen in (b). This indicates the non-Newtonian fluid behaviour. The models
describing such non-parabolic profiles exist (see, e.g., reference~\cite{Fedosov2010}) and involve
an analogue for the viscosity and a number of additional parameters. We found, however,
a numerical fitting to these forms impractical. The set of rules defining the behaviour of the
particles at walls, as imposed in our simulation, leads here to the slip boundary conditions.
The discontinuity of the velocity profile at the wall boundary is clearly visible in
figure~\ref{vx_fit}~(c) and is a characteristic feature of the flows of polymeric fluids.
In figure~\ref{vx_fit}~(d), we compare two velocity profiles of the flows obtained with
applying equation~(\ref{Gauss_distr})--(\ref{gauss_norm}), i.e., the Gaussian distribution of the bulk force,
and a uniform distribution of the bulk force. We note that even for such an extremely large value
of the bulk force, the profiles are practically identical. We expect that for very long polymers, the Gaussian
distribution of the bulk force would prove beneficiary and could lead to better stability
of the integration of the equations of motion.

An alternative route is to concentrate on the details of the microstructure of the oligomer flow, because these
must be responsible for its non-Newtonian behaviour. In particular, comparing to the case of a simple
fluid, oligomers have additional conformational degrees of freedom which will affect their flow properties.
Therefore, we build the profiles for the average shape anisotropy and the molecular orientation
for the oligomers in a flow. The components of the gyration tensor
\begin{equation}
\label{eq:gyr_tensor}
G_{\alpha,\beta}=\frac{1}{L\idx{o}}\sum\limits_{i=1}^{L\idx{o}}(r_{i,\alpha}-R_{\alpha})
(r_{i,\beta}-R_{\beta})
\end{equation}
are evaluated for each oligomer of length $L\idx{o}$ at a given time instance $t$.
Here, $\alpha,\beta$ denote the Cartesian axes, $r_{i,\alpha}$ are the coordinates of
$i$th monomer, and $R_{\alpha}$ are the coordinates for the center of mass of
the oligomer. In the equivalent ellipsoid representation, the eigenvalues
$\lambda_1>\lambda_2>\lambda_3$ of this tensor provide the squared lengths of its semiaxes,
whereas the respective eigenvectors $\vvec{u}_1$, $\vvec{u}_2$ and $\vvec{u}_3$~---
the orientation of these axes in space.

The shape anisotropy of an individual oligomer can be defined as
\begin{equation}\label{k2}
\kappa^2=\frac{3}{2}
\frac{\lambda_1^2+\lambda_2^2+\lambda_3^2}
     {\left[\lambda_1+\lambda_2+\lambda_3\right]^2}
    -\frac{1}{2}.
\end{equation}
It is zero for a spherically symmetric body, where $\lambda_1=\lambda_2=\lambda_3>0$
and is equal to $1$ for an infinitely long thin rod, where $\lambda_1>0$, $\lambda_2=\lambda_3=0$.
The average profile is built for the shape anisotropy in a steady state. It is obtained by first
binning the system in $Z$-axis and averaging $\kappa^2$ for individual oligomers found in each bin.
Then, time averaging within the steady state is performed.

The orientation of each oligomer in space is defined by that for the longest axis of its equivalent ellipsoid.
The latter is characterised by the eigenvector $\vvec{u}_1$ associated
with the largest eigenvalue $\lambda_1$. The level of alignment of the oligomer along the flow axis $X$
can be characterised by the order parameter:
\begin{equation}\label{Sx}
S_x=P_2(\vvec{u}_1\cdot \vhat{x}),
\end{equation}
where  $\vhat{x}$ is defined in equation~(\ref{FF}) and $P_2(x)$ is the second Legendre polynomial. The alignment profile is built then in a steady state
by averaging $S_x$ in each bin and then performing time averaging.
It is obvious that both $\kappa^2$ and $S_x$ can be defined for the case $L\idx{o}>1$ only.

\begin{figure}[!t]
\centering
\includegraphics[width=0.95\textwidth]{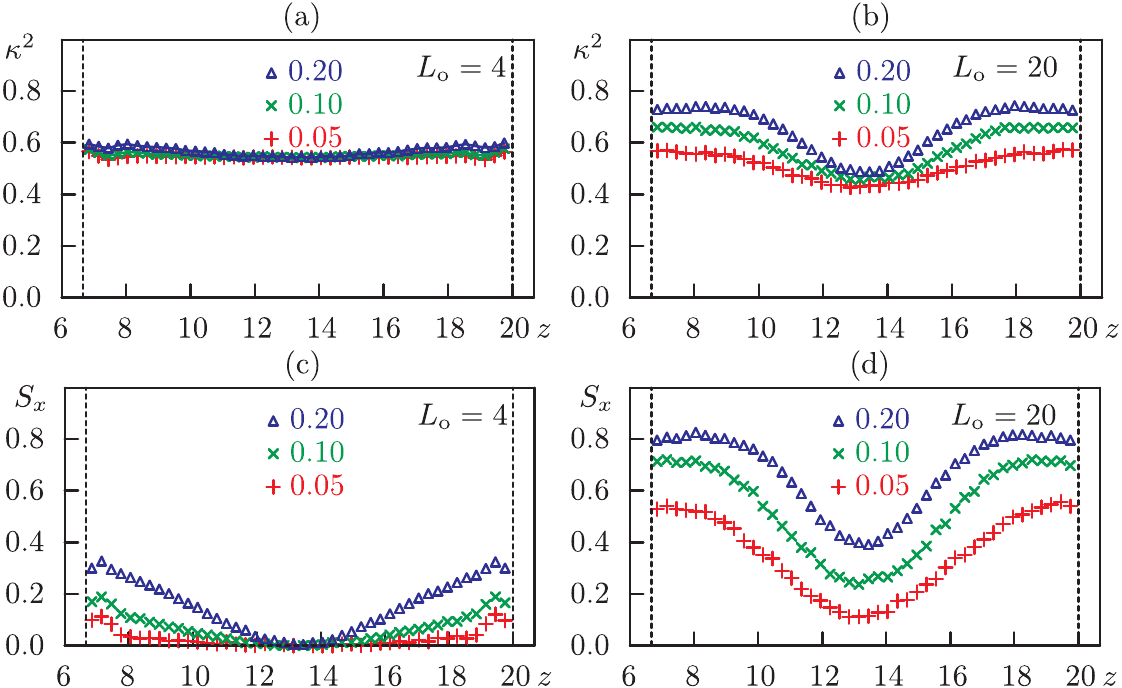}
\caption{\label{anis_ordx_olig}  (Color online) Profiles for average shape anisotropy [(a) and (b)] and oligomer alignment
[(c) and (d)]. Respective oligomer length $L\idx{o}$ and the magnitude of the flow force are indicated
in each plot.}
\end{figure}

Average shape anisotropy and oligomer alignment profiles are shown in figure~\ref{anis_ordx_olig}
for the shortest $L\idx{o}=4$ and longest $L\idx{o}=20$ oligomer considered in this study.
The case $L\idx{o}=4$ is characterised by flat anisotropy profile with the value $\kappa^2\approx 0.6$
independent of the magnitude of the flow force [see, frame (a)]. One can conclude that for the oligomer being
this short, the flow does not change its shape (at least for the flow force magnitude range used here).
For the longest oligomer, $L\idx{o}=20$ [frame (b)], the average value of $\kappa^2$ over the profile is close to that for
$L\idx{o}=4$, but the profile exhibits distinct shoulders with higher $\kappa^2$ values near both channel edges
and a well in its center. The channel edges, therefore, promote a stronger anisotropy for the adjacent oligomers,
presumably due to entropic effects.

The alignment profile for the shortest oligomer [frame (c)] indicates no orientational order in the middle
of the channel ($S_x\approx 0$), whereas a relatively weak alignment is observed near the channel edges,
which rises to $S_x\sim 0.35$ with the increase of the flow force magnitude $f$. For the longest oligomer [frame (d)],
the alignment profile has a cosine-like shape which moves up almost unchanged with an increase of $f$. It is
non-zero in the middle of a channel for all $f$ being considered. Therefore, at least for longer oligomers, $L\idx{o}>4$,
there is a variation of the oligomer shape and alignment across the channel: the molecules are found to be
much more elongated and aligned near the edges as compared to the middle part.

The flow-induced deformation of the polymer molecules renders their shape to be more similar to
liquid crystals. The effect is detected for longer oligomers $L\idx{o}>4$ and stronger flows,
where the effective length-to-breadth ratio of oligomer exceeds a certain threshold. Similar effect
is well known for the systems of anisotropic hard bodies, where the orientationally ordered phases
are also found above certain threshold length-to-breadth ratio \cite{Hard_part-a,Hard_part-b,Hard_part-c,Hard_part-d}.
Using this liquid crystal analogy, we recall the results obtained by Mazza {et al.} \cite{Mazza2010,Mazza2011}
reporting the high self-diffusivity of the Gay-Berne-Kihara fluid along the director in the ``supernematic'' phase.
Following these findings, one expects an essential reduction of the friction between the aligned oligomers near the
channel edges, as compared to that in the central part. Larger friction between oligomers in the middle
of a channel is seen as the reason for the suppression of the velocity profile here and, as a result,
its non-parabolic, bell-like shape [cf. figure~\ref{vx_fit}~(b)], and the appearance of the slip boundary
condition [cf. figure~\ref{vx_fit}~(c)].

\clearpage

\section{\label{III}Flow of oligomeric fluid through a channel with polymer modified walls}

\begin{figure}[!b]
\centering
\includegraphics[width=8.5cm]{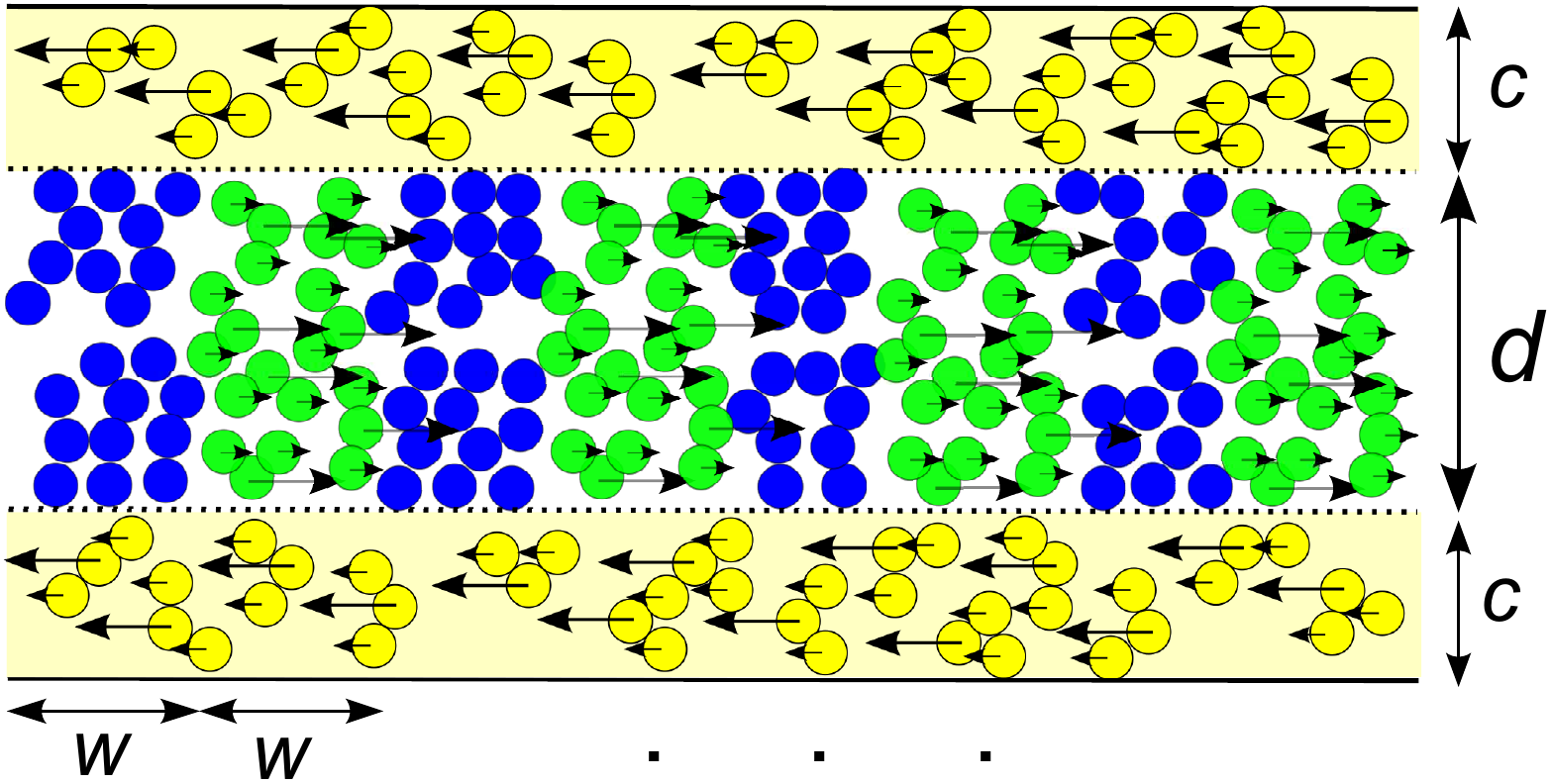}
\caption{\label{geom_brush_olig} (Color online) Extension of the simulation box geometry
  of figure~\ref{geom_nobr_olig} to the case of a slit-like pore with its internal walls modified
  by stripes of polymer brushes (displayed in blue). The stripes are of width $w$ and
  are periodic along the $X$-axis (direction of flow). Pore interior is filled with oligomeric
  fluid (displayed in green). The flow force is applied to the oligomers only.}
\end{figure}

We turn now to the case when the pore walls are modified by polymer brushes arranged in the form of stripes
(see, figure~\ref{geom_brush_olig}). Each chain of a brush is of length $L=20$ beads of type $A$, the pore interior is
filled with the oligomer fluid of the length $L\idx{o}$ beads of type $B$, the contraflow regions contain
an oligomer fluid of the length $L\idx{o}$ beads of type $A$. The difference between the bead types is in the
value of the repulsion amplitude $a$ in equation~(\ref{FC}) being set to $a_{AA}=a_{BB}=25$ and $a_{AB}=40$ for
the interaction of similar and dissimilar beads, respectively. Therefore, the oligomer acts as a bad solvent for the
brush. The good solvent case, $a_{AA}=a_{BB}=a_{AB}=25$, is briefly discussed in the end of this section.

The equilibrium properties of the setup depicted in figure~\ref{geom_brush_olig} for the case of $L\idx{o}=1$
and no contraflow regions are studied in detail in reference~\cite{ISP1,ISP2}. Equilibrium morphology was
found to depend on the parameters $d$ and $w$, and is formed as a  result
of an interplay between the enthalpy and the entropy of the system. In particular, at
small $w\ll L$ and any $d$, the adjacent brush stripes belonging to the same wall merge
and form a homogeneous ``coat'' on the wall resulting in the lamellar morphology.
In this case, the chains are stretched and aligned along the $X$-axis. With an
increase of $w$, the adjacent brush stripes are incapable of merging any more. Instead, they
either stay separately (at relatively large $d\sim L$) or merge across the pore
with their counterparts grafted to the opposite wall to form a pillar phase (at
small enough $d<L$). In this case, the brush chains are stretched and aligned in $Z$
direction. This demonstrates a strong correlation between the alignment direction of brush chains
and the topology of the equilibrium morphology. Therefore, it looks plausible that
the change of the alignment of the brush chain by means of an external stimulus could
result in a morphology change in the system.

\begin{figure}[!t]
\centering
\includegraphics[width=8cm]{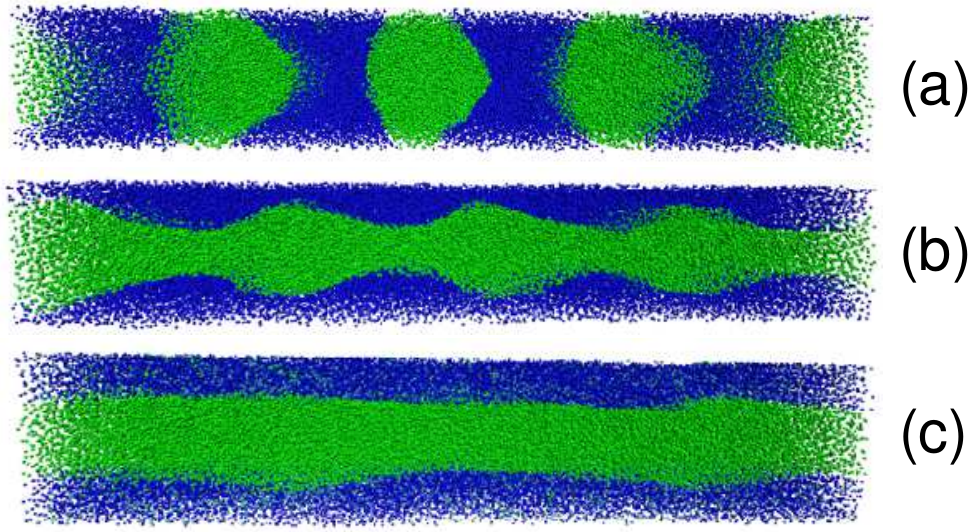}
\caption{\label{snap_d13_w10}  (Color online) Sequence of snapshots showing the flow-induced
transitions from pillar (a) through modulated lamellar (b) into
flatten lamellar (c) morphology. System geometry: $d=13.333$, $w=10$, $c=4$,
$L\idx{o}=1$, the flow force magnitude is $f=0.02$, $0.1$, and $0.4$ for (a), (b) and (c),
respectively. Colours follow these in figure~\ref{geom_brush_olig}, contraflow regions
not shown.}
\end{figure}

\begin{figure}[!t]
\centering
\includegraphics[width=0.95\textwidth]{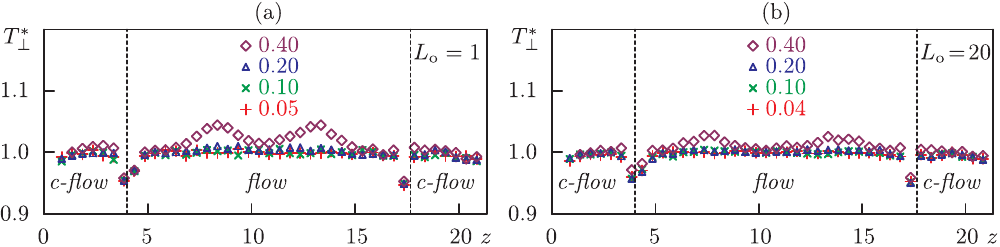}
\caption{\label{t_conserv_br}  (Color online) $T_{\perp}^*$ profile at various amplitudes of the bulk
force $f$ indicated in the figure. Geometry $d=13.333$, $w=10$, $c=4$
is considered with polymer-modified pore boundaries. (a) simple liquid, $L\idx{o}=1$;
(b) longest oligomer simulated, $L\idx{o}=20$.}
\end{figure}

This is the case, indeed, when a flow force above certain threshold value is applied
to the fluid in the pore. Let us consider first the visual representation of morphology
changes in the form of a snapshot sequence. The case of $d=13.333$, $w=10$, $c=4$,
at various values of the force $f$ is presented in figure~\ref{snap_d13_w10}.
For this geometry of a pore, a  stable pillar phase is observed when no or weak flow
force is applied [cf. reference~\cite{ISP1} and figure~\ref{snap_d13_w10}~(a)].
With an increase of $f$ above the threshold value of $f\approx 0.04-0.06$, the pillars
break and the morphology switches to the modulated lamellar morphology [see
figure~\ref{snap_d13_w10}~(b)]. The layers, formed of brush chains bent along the
flow, gradually flatten as $f$ increases further, as shown in figure~\ref{snap_d13_w10}~(c). One should remark that a perfect stationary lamellar morphology is also aided
by a microphase separation between the $A$ beads of tethered chains and $B$ beads of
the flowing oligomer.

\begin{figure}[!b]
\centering
\includegraphics[width=0.95\textwidth]{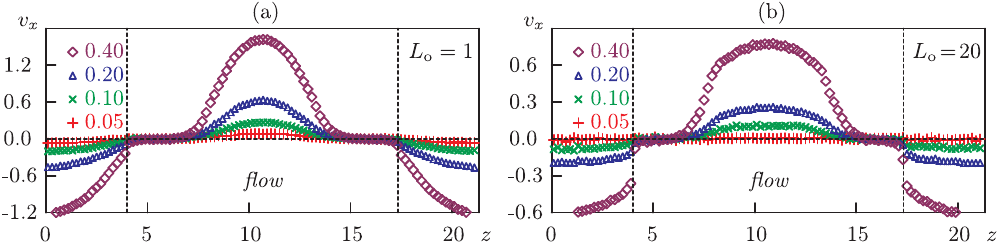}
\caption{\label{vx_fit_br}  (Color online) Profile for velocity component along the flow direction, $v_x$,
evaluated bead-wise at several values of $f$ (indicated in the figure) for
the setup depicted in figure~\ref{geom_brush_olig}, the other parameters are the same
as in figure~\ref{t_conserv_br}.}
\end{figure}

Let us check the quality of temperature conservation, similarly to the analysis performed
in section~\ref{II} for the pore with flat boundaries. As follows from figure~\ref{t_conserv_br}
(a), maximum deviation of the temperature profile from the required value $1$ does not exceed
$4\%$ if the force amplitude is restricted to $f\leqslant 0.4$ for both cases of $L\idx{o}=1$ and
$L\idx{o}=20$.
It is worth mentioning that the maximum usable value for $f=0.4$ here is twice as large
as its counterpart for the case of flat boundaries, see figure~\ref{t_conserv}. This relation
can be attributed to the fact that the total amount of the force applied inside a pore
with polymer modified boundaries (figure~\ref{t_conserv_br}) is twice less compared
to the case of the pore with flat boundaries (figure~\ref{t_conserv}). This is so due to
the fact that no force is applied to the polymer brush beads (which are half of all the beads in the system).

The average profiles for the velocity component along the flow direction, $v_x$ are
shown in figure~\ref{vx_fit_br} at various force amplitudes $f=0.05$, $0.1$, $0.2$
and $0.4$ for the cases of $L\idx{o}=1$ and $L\idx{o}=20$. Comparing these profiles
with their counterparts for the case of flat internal walls (figure~\ref{vx_fit}), one can make
the following observations. Firstly, at $f\geqslant 0.1$, the profiles exhibit two ``shoulders'' near
each internal wall which are characterized by zero values for $v_x$. These are, obviously,
the regions occupied by the polymer brush which ``envelopes'' the internal walls
(see, figure~\ref{snap_d13_w10}). The flow is completely suppressed within these layers,
rendering the walls thicker and reducing the pore size accessible to the flow.
As a consequence, the maxima for $v_x$ decrease  compared to the case of flat walls.
Secondly, the shape of the central part of each velocity profile follows closely their counterparts in
figure~\ref{vx_fit}. It is of parabolic shape for $L\idx{o}=1$ and bell-like for $L\idx{o}=20$,
suggesting qualitative similarities between both flows.

\begin{figure}[!t]
\centering
\includegraphics[width=0.95\textwidth]{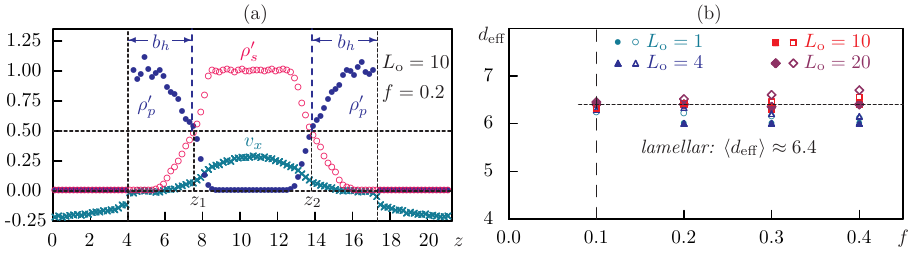}
\caption{\label{d_eff_schema}  (Color online) (a) Schema explaining the estimates for the effective pore size $d\idx{eff}$ in
stationary lamellar morphology for $L\idx{o}=10$, and $f=0.2$,
$\rho'_p$ and $\rho'_s$ are reduced density profiles for brush and oligomer beads, respectively.
$b_h$ is brush thickness (\ref{eq:brush_thickness}), $z_1$ and $z_2$ are the intersection points
of $\rho'_p$ and $\rho'_s$. The velocity profile $v_x$ is also shown (not in scale).
(b) $d\idx{eff}$ at various $L\idx{o}$ and $f$. Solid legends: estimates from the intersection
points $z_1$ and $z_2$, open legends: esimates from the brush height $b_h$.
The dashed vertical line marks the formation of the lamellar morphology.}
\end{figure}

This interpretation brings up the possibility to treat a fluid flow within a stationary
lamellar morphology similarly to the case of the pore with flat boundaries, discussed
in section~\ref{II}, except for the smaller effective pore size $d\idx{eff}$ \cite{Pastorino2009}.
To evaluate the latter, one can use the expression for an average brush thickness
\begin{equation}
\label{eq:brush_thickness}
b_h=2\frac{\int \tilde{z}\rho_p(\tilde{z})\rd\tilde{z}}{\int \rho_p(\tilde{z})\rd\tilde{z}},
\end{equation}
where $\rho_p(\tilde{z})$ is the density profile of the beads that belong to the tethered
chains and $\tilde{z}$ is the distance from the nearest pore boundary along the
$Z$-axis. In this case, one obtains $d\idx{eff}=d-2b_h$. Alternatively, the effective
pore size can be estimated as the distance between the intersection points $z_1$
and $z_2$ for $\rho'_p=\rho_p/\rho$ and $\rho'_s=\rho_s/\rho$, the reduced
density profiles for the polymer and the flowing oligomer beads, respectively. This is
illustrated in figure~\ref{d_eff_schema}~(a). It is evident that, for
this particular case, both estimates for $d\idx{eff}$ are extremely close. To check how
this observation holds for other oligomer lengths $L\idx{o}$ and flow forces $f$, we
performed both types of estimates for $d\idx{eff}$ in each case. The results are
presented in figure~\ref{d_eff_schema}~(b), where the estimates for $d\idx{eff}$ made from
the intersection points $z_1$ and $z_2$ are presented via solid legends, whereas
the estimates performed via the evaluation of $b_h$ are shown via open symbols.
One can make several conclusions from this plot. First, the value of $d\idx{eff}$,
evaluated by both approaches, are similar to each other for $f\geqslant 0.1$.
This threshold correlates well with the value of $f$, at which the stationary
lamellar morphology is formed (marked with the dashed line in figure~\ref{d_eff_schema}).
While $d\idx{eff}$ can be also calculated at smaller values of $f$,
these results would carry no physical significance due to the pillar morphology.
Second, the difference between the values for $d\idx{eff}$ estimated by means
of two alternative methods at the same $L\idx{o}$ and at the same $f$,
does not exceed $4\%$. Therefore, either of the estimates for $d\idx{eff}$ can be used.
Third, there is a trend for an increase of $d\idx{eff}$ with the growth of oligomer length
$L\idx{o}$, although it is rather modest. For example, for the case of $f=0.4$ the value of
$d\idx{eff}$ for oligomer length of $L\idx{o}=20$ is only $10\%$ higher than its
counterparts for $L\idx{o}=1$ and $4$, and this increase is of the order of the error
in the estimates of $d\idx{eff}$ mentioned above.
A slight increase of the brush height with an increasing flow can be attributed to the fact
that there is some residual flow of oligomers inside the brush.
As the flow increases, the flow-induced elongation of the oligomers leads to an increase
of their effective size and this will lead to a slight increase of the brush height.

As was discussed in section~\ref{II}, flat walls of the setup depicted in figure~\ref{geom_nobr_olig}
act as effective ``stretchers'' and ``aligners'' for the adjacent oligomer molecules, which results in
characteristic profiles for $\kappa^2$ and $S_x$ shown in figure~\ref{anis_ordx_olig}.
It is, therefore, of interest to see whether or not the effective walls formed by a flattened polymer brush,
as pictured in figure~\ref{snap_d13_w10}~(b) and (c), have a similar impact on the adjacent oligomer molecules.
We examine the aligning capabilities of such flattened brushes more in detail, considering both cases
of bad and good oligomer solvent. For the former case, the repulsion parameter $a$ in equation~(\ref{FC})
is set to $a_{AB}=40$ for the interaction between oligomer and brush monomers, whereas for the latter
case we set $a_{AB}=25$. The repulsion parameter between similar beads is equal to $a_{AA}=a_{BB}=25$
in both cases.

\begin{figure}[!t]
\centering
\includegraphics[width=0.85\textwidth]{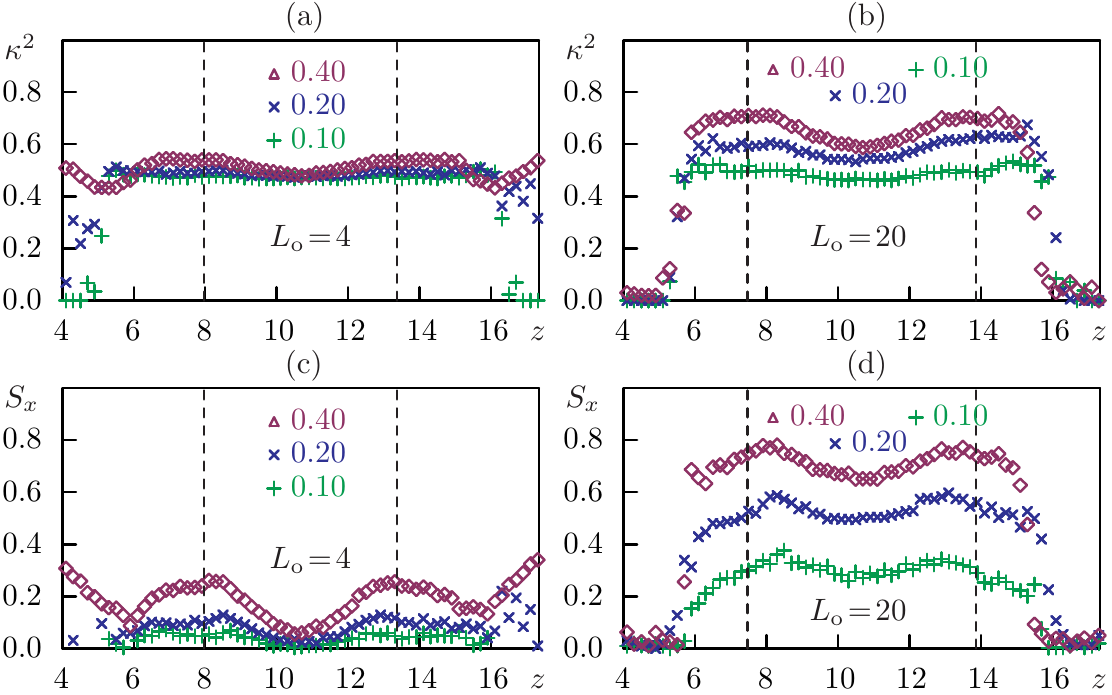}
\caption{\label{anis_ordx_olig_br}  (Color online) Profiles of the average shape anisotropy [(a) and (b)]
and oligomer alignment [(c) and (d)] for the geometry depicted in figure~\ref{geom_brush_olig}
and the case of bad oligomer solvent. Respective oligomer length $L\idx{o}$ and the magnitude
of the flow force are indicated in each plot.}
\end{figure}

\begin{figure}[!b]
\centering
\includegraphics[width=0.85\textwidth]{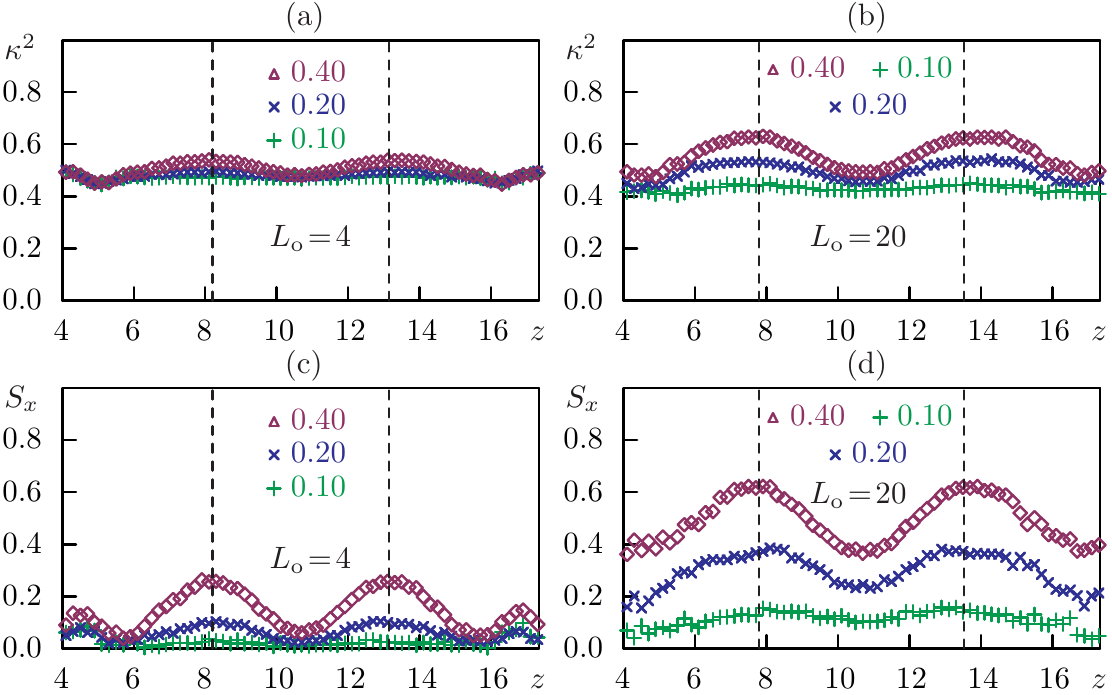}
\caption{\label{anis_ordx_olig_br_goodsolv}  (Color online) The same as in figure~\ref{anis_ordx_olig_br}
but for the case of good oligomer solvent.}
\end{figure}

For the bad solvent case, the flattened brush and oligomer flow are strongly demixed with oligomers
being expelled from the brush-rich regions. The profiles for $\kappa^2$ and $S_x$ are shown in
figure~\ref{anis_ordx_olig_br} for the oligomer lengths of $L_{\mathrm{o}}=4$ and
$L_{\mathrm{o}}=20$. Here, we make use of our estimates for the effective pore size, $d\idx{eff}$,
indicated on each plot by vertical dashed lines. If restricted to this region, then
the profiles depicted in figure~\ref{anis_ordx_olig_br} are extremely close
to their respective counterparts for the case of flat walls shown in figure~\ref{geom_nobr_olig}
in both shape and absolute values, save for being ``squeezed'' into the smaller pore
size $d\idx{eff}$. This indicates that the existing roughness of the flattened polymer wall
does not reduce its impact on the adjacent oligomer molecules.

For the good solvent case, the brush and oligomer mix well if no flow force is applied,
but a flow-driven lamellarization of the system takes place at about $f=0.2$, similarly
to the bad solvent case. The profiles for $\kappa^2$ and $S_x$ are shown in
figure~\ref{anis_ordx_olig_br_goodsolv} for the oligomer lengths of $L_{\mathrm{o}}=4$ and
$L_{\mathrm{o}}=20$. The estimated effective pore size $d\idx{eff}$ is of the same order
but a fraction smaller than that for the bad solvent case. This is indicated in
figure~\ref{anis_ordx_olig_br_goodsolv} by vertical dashed lines. One should remark that
despite the strong alignment of the polymer brush and good mixing between the brush and the oligomers,
the latter are found less elongated and less aligned along a flow compared to the bad solvent
case shown in figure~\ref{anis_ordx_olig_br}. The respective curves for $\kappa^2$ are lower
by about $0.1$ compared to their counterparts for the bad solvent case, whereas these
for $S_x$ are about $0.2$ lower. One should attribute this to the fact that the flattening
of the brush in the case of a good solvent requires a higher flow force compared to the case
of a bad solvent. In the latter case, lamellarization is also aided by the microphase
separation between the brush and the oligomer fluid.

Despite these quantitative differences, the qualitative picture emerging for both cases of the bad
and good solvent is essentially the same. Namely, at a certain value of the flow force $f$, the stationary lamellar
phase is formed with the flowing oligomer occupying the center part of the pore.
The oligomer is found essentially elongated and aligned along the flow near the walls of
this channel and much less in the center of the pore. This effect is detected for the oligomer
lengths $L_{\mathrm{o}}>4$ and, due to its impact on the distribution of the local friction across
the pore, affects the behaviour of the fluid turning it into a non-Newtonian one.

\section{\label{IV}Conclusions}

In this paper, we developed the simulation approach, which allows one to simulate the pressure-driven  flow
in the pore with flat and polymer-modified walls. It combines the earlier ideas of fluid-like walls
and reverse flow. The former enables to avoid highly structured solid walls that usually
lead to near-wall artefacts. The latter introduces friction between oppositely flowing streams which
makes it possible to conserve the total momenta and keep the temperature constant. Our system geometry contains
the central main pore ``enveloped'' by two fluid-like pores on each side. The flow force is introduced in
the main pore and oppositely directed contra-flow of the same magnitude~--- in both fluid-like walls.
Simulation of the oligomer flow through the pore with flat walls is used as a check for the credibility
of the method and reproduction of the hydrodynamic boundary conditions.
Good thermostatting of the system is achieved when the flow force magnitude does not
exceed a certain threshold. For the case of the oligomer length $L_{\mathrm{o}}>4$, we found the
molecules adjacent to the central pore boundaries essentially stretched and aligned along the flow,
whereas their shape is more spherical and less aligned in the middle of a pore. This provides the
basis for variation of the local friction across the pore and, as a consequence, the non-parabolicity
of the velocity profile for the oligomer fluid and the slip boundary condition.

The case of polymer-modified walls is also considered when the polymer brush has the form of
stripes arranged perpendicularly to the flow direction. In this case, at a certain threshold value of
the flow force, one observes the pillar-to-lamellar transition induced by the flow which leads to
the brush enveloping the pore wall with a relatively smooth layer. At higher flow rates,
the flow of oligomer is similar to the case of flat walls, although for the
narrower effective pore size. The latter is estimated both from the intersection of density profiles
for the brush and the flowing oligomer and from the integral equation for the average brush thickness.
The effect of local stretching and alignment of oligomers near the walls of the effective pore
is detected the same as for the case of flat walls.

The method can be extended to more complex systems, namely: the flow of mixtures and their
flow-induced separation; the flow of amphiphilic molecules; the flow of complex macromolecules
or their solutions. Combined with the fine-tunable structure of the brush, this opens up a
possibility to study various problems of transport of oligo- and macromolecules through
a complex structured environment.

\clearpage

\section*{Acknowledgements}

This work was supported by the EU under IRSES Project STCSCMBS 268498.


\ukrainianpart

\title{Потік олігомерного флюїду в нано-каналі із комплексною
структурою під впливом зовнішнього тиску. \\ Дослідження методом
дисипативної динаміки
}
\author{Я.М. Ільницький\refaddr{label1}, П. Брик\refaddr{label2}, А. Патрикєєв\refaddr{label2}}
\addresses{
\addr{label1} Інститут фізики конденсованих систем НАН України, вул. І.~Свєнціцького, 1, 79011 Львів, Україна
\addr{label2} Відділ моделювання фізико-хімічних процесів, Університет Марії Кюрі-Склодовської, \\ 20--031 Люблін, Польща
}
\makeukrtitle
\begin{abstract}
Розвинуто симуляційний метод, спрямований на моделювання потоку у порі із гладкими та полімер-модифікованими стінками.
Підхід грунтується на методі дисипативної динаміки і реалізує
ідеї ``рідинних стінок'' та реверсних потоків.
Як тест розглянуто потік олігомерів крізь пору із гладкими стінками
і продемонстровано добре термостатування системи при застосуванні цього методу. Отримано неоднорідності як форми плинних олігомерів, так і ступеня їх вирівнювання вздовж пори, які призводять до
непараболічних профілів швидкостей. Метод застосовано до наноканалу,
декорованого смугами полімерних щіток, розташованих перпендикулярно до напрямку потоку. При певному граничному значенні сили потоку зафіксовано перехід із стовпцевої до ламеларної морфології, який призводить до обгортання стінок пори гладким шаром із полімерних щіток. При вищих швидкостях потік олігомерів набуває властивостей аналогічних до випадку гладких стінок, але із вужчим ефективним розміром пори. Спостережено розтяг і вирівнювання полімерних молекул вздовж напрямку потоку поблизу стінок пори.
\keywords Пуазейлевий потік, полімерна щітка, олігомери, метод дисипативної динаміки
\end{abstract}

\end{document}